\documentclass[12pt]{article}
\usepackage{latexsym}
\usepackage{amsmath}

\renewcommand{\epsilon}{\varepsilon}

\renewcommand{\baselinestretch}{1.7}

\usepackage{amsmath,bm}
\usepackage{amssymb}
\usepackage{xcolor}
\usepackage{color}

\usepackage{amsthm}
\usepackage[ansinew]{inputenc}
\usepackage{graphicx}
\usepackage{epsfig}
\usepackage{dsfont}
\usepackage{bbm}
\usepackage{subfig}
\usepackage{url}
\usepackage{placeins}
\usepackage{epstopdf}
\usepackage{enumitem}

\usepackage[authoryear]{natbib}

\newtheorem{satz}{Theorem}[section]
\newtheorem{theorem}{Theorem}[section]

\newtheorem{remark}[satz]{Remark}

\newtheorem{algorithm}[satz]{Algorithm}

\def\3{\ss}

\newcommand{\bea}{\begin{eqnarray*}}
\newcommand{\eea}{\end{eqnarray*}}
\newcommand{\be}{\begin{eqnarray}}
\newcommand{\ee}{\end{eqnarray}}

\newcommand{\ba}{\begin{array}}
\newcommand{\ea}{\end{array}}

\def\3{\ss}

\newcommand{\dn}{\stackrel{\cal D}{\longrightarrow}}

\usepackage[authoryear]{natbib}

\begin{document}
\def\spacingset#1{\renewcommand{\baselinestretch}%
{#1}\small\normalsize} \spacingset{1}

\title{{\bf Equivalence  of regression curves sharing common parameters}}

\author
{Kathrin M\"ollenhoff$^{1}$,
	Frank Bretz$^{2}$  ,
	Holger Dette$^{1}$ \\
	$^{1}$	Department of Mathematics, Ruhr-Universit\"at Bochum, Germany,\\
	$^{2}$Novartis Pharma AG, CH-4002 Basel, Switzerland
}

\pdfminorversion=4
 \maketitle

\begin{abstract}
In clinical trials the comparison of two different populations is a frequently addressed problem. Non-linear (parametric) regression models are commonly used to describe the relationship between covariates as the dose and a response variable in the two groups. In some situations it is reasonable to assume some model parameters to be the same, for instance the placebo effect or the maximum treatment effect. 
In this paper we develop a (parametric) bootstrap test to establish the similarity  of two regression curves sharing  some common parameters. We show by theoretical arguments and by
means of a simulation study that the new  test controls its level and achieves a reasonable power. Moreover,  it is demonstrated that 
under the assumption of common parameters   a considerable more powerful test  can be constructed compared 
to the test which does not use this assumption. Finally, we illustrate potential applications of the new  methodology by a clinical trial example. 
\end{abstract}

\parindent 0cm

\spacingset{1.2}
\section{Introduction} \label{sec1}
\def\theequation{1.\arabic{equation}}
\setcounter{equation}{0}

Regression models are commonly used to describe the relationship between multiple covariates and a response variable. In certain applications, more than one regression model is available, such as when assessing the relationship between the covariates and the response variables in more than one population (e.g. in males and females). It is then often of interest to demonstrate the equivalence of the regression curves: If equivalence can be claimed, conclusions can be drawn from the pooled sample and a single regression model is sufficient to describe the data. This can be achieved by testing a suitable null hypothesis that the distance between the regression curves (measured in an appropriate sense) is smaller than a pre-specified equivalence margin at a controlled Type I error rate. Note that the problem of equivalence testing, as considered in this paper, is conceptually different from the more frequent problem of testing for equality of curves and is much less studied in the literature due to methodological difficulties.

The problem of testing for equality  of regression models has been intensively discussed in the nonparametric context  and we refer 
to the recent work of \cite{Feng2015}, whichcontains a rather comprehensive list of references.
In applied regression analysis, however, parametric models are usually preferred to a purely nonparametric approach as they admit a direct interpretation of the observed effects in terms of the model parameters. In addition, the available information of the observations is increased by applying more efficient estimation or test procedures, provided that the assumed model is valid.
Despite its importance, the problem of establishing equivalence of two parametric regression models while controlling the Type I error rate has only recently found attention in the literature. Using the intersection-union test device from \cite{berger1982}, \cite{liubrehaywynn2009} investigated the assessment of non-superiority, non-inferiority and equivalence when comparing two regression models over a restricted covariate region. Building upon this work, \cite{gsteiger2011} derived equivalence tests based on simultaneous confidence bands for nonlinear regression models, with application to population pharmacokinetic analyses. Likewise, \cite{bretzmoelldette2016} assessed the similarity of dose response curves in two non-overlapping subgroups of patients. Alternatively, \cite{detmolvolbre2015} suggested directly estimating the distance between the regression curves and using a non-standard bootstrap test to decide for equivalence of the two curves if the estimate is less than a certain threshold. Expanding this approach, \cite{moellenhoff2018} assessed the comparability of drug dissolution profiles via maximum deviation, whereas \cite{hoffelder2018} demonstrated the equivalence of dissolution profiles using the Mahalanobis distance; see also \cite{collignon2018}.

In these papers, the authors assumed that the regression models have different parameters and can therefore be evaluated separately. In some applications, however, this assumption
cannot be justified and  it is more reasonable to assume  that the regression models may have  some common parameters. The total number of parameters to estimate is then reduced to the common and remaining parameters of each model, affecting the asymptotic behavior of the estimators. Consider, for example, the Phase II dose finding trial for a weight loss drug described in \cite{bretzmoelldette2016}. This trial aimed at comparing the dose response relationship for two regimens administered to patients suffering from overweight or obesity: Three doses each for once daily (o.d.) and twice daily (b.i.d.) use of the medication, and placebo. It is reasonable to assume that the placebo response is the same under both the o.d. and the b.i.d. regimen. Since the regression models typically used for dose response modeling contain a parameter for the placebo response (\cite{pinheiro2006analysis}), they will thus share this common parameter for both the o.d. and the b.i.d. regimen. In some instances, it might even be reasonable to assume  that the maximum efficacy for high doses is similar
in both groups. Moreover, clinical trial sponsors may even decide to use the same placebo group for logistical reasons. The response of each patient on placebo is then used twice in the estimation 
of the o.d. and b.i.d. dose response models, further complicating the statistical problem.

In this paper, we investigate the equivalence of two parametric regression curves that share common parameters. In Section~\ref{sec2} we first introduce the regression models to be estimated under the assumption of common parameters. We then develop  a non-standard bootstrap test which performs the resampling under the constraints of the interval hypotheses implied by the equivalence test problem.
The new tests improves the  procedure proposed in \cite{detmolvolbre2015} using the additional information of common parameter in both groups.
We also discuss testing the equivalence of model parameters to assess whether the assumption of common parameters is plausible. In Section~\ref{sec3} we investigate the finite sample properties of the proposed bootstrap test proposed in terms of power and size. In Section~\ref{sec4} we illustrate the methods using a multi-regional clinical trial example where it is conceivable that the placebo and maximum treatment responses are the same across geographic regions but the onset of treatment differs due to intrinsic and extrinsic factors (\cite{malinowski2008}; \cite{ich2017}). Technical details and proofs are deferred to an appendix.

\section{Methodology} \label{sec2}
\def\theequation{2.\arabic{equation}}
\setcounter{equation}{0}

\subsection{Models with common parameters}\label{sec2.1}

Let
\be \label{mod1a}
Y_{\ell ,i,j} = m_\ell (d_{\ell ,i},\beta_\ell )+\eta_{\ell ,i,j}~,~j=1,\ldots , n_{\ell ,i},~i=1, \ldots, k_\ell, 
\ee
denote the observed response of the $j$th subject at the $i$th dose level $d_{\ell,i}$ under the $\ell$th dose response model $m_\ell$, 
where $\ell =1,2$  denotes the index of the two groups under consideration.
We assume that the (non-linear) regression model $m_\ell$ is parametrized through a $p_\ell$-dimensional vector $\beta_\ell$,  
$\ell =1,2$.
Note that the regression models $m_1$ and $m_2$ may be different. Likewise, the parameters $\beta_1$ and $\beta_2$ may be different even if $m_1 = m_2$.
We further assume that the error terms $\eta_{\ell,i,j}$ are independent and identically distributed with expectation $0$ and variance $\sigma_\ell^2$. 
The dose levels $d_{\ell,i}$ may be different in  both groups but they are attained on the same (restricted) covariate region $\cal D$.
In this paper $\cal D$ is assumed to be the dose range, although the results can be generalized to include other covariates.
Further, $n_\ell =\sum_{i=1}^{k_\ell}n_{\ell ,i}$ denotes the sample size in group $\ell$ where we assume $n_{\ell,i}$ observations in the $i$th dose level ($i=1, \ldots, k_\ell,\ \ell=1,2)$.
The sample sizes $n_\ell$ can be unequal and the total number of observations is denoted by $n=n_1+n_2$.

In this paper we consider the situation, where  the regression models  have some common parameters. More precisely, we  assume 
without loss of generality that these parameters are given by the first $p'$ model parameters  of  the parameter $\beta_{\ell}$ in model \eqref{mod1a}
that is  
\be\label{common_parameters}
\beta_\ell=(\beta_0,\tilde \beta_\ell) \in\mathbb{R}^{p_\ell}  , ~~\ell=1,2 ~,
\ee
where  $\beta_0\in\mathbb{R}^{p'}$ denotes the vector of  common parameters in  both regression  models and $\tilde \beta_1$ and $\tilde \beta_2$ denote the remaining parameters in the
models $m_{1}$ and $m_{2}$, respectively,
which do not necessarily coincide. The case where the models $m_{1}$ and $m_{2}$ do not share any common parameters is included and corresponds to
$\beta_\ell=\tilde \beta_\ell$  for $ \ell=1,2$ (that is $p^{\prime} =0$).  As  a consequence the  $p_1+p_2-p'$-dimensional  vector of all   parameters of  the regression functions in  
model \eqref{mod1a} 
under the assumption \eqref{common_parameters} is given by $\beta=(\beta_0,\tilde \beta_1,\tilde \beta_2)$. Throughout this paper we assume
that  $\beta \in B $ where $B \subset  \mathbb{R}^{p_1+p_2-p'}$ is a compact set.

These parameters are  now  estimated by  least squares using the  combined sample $\{ Y_{\ell ,i,j} : ~j=1,\ldots , n_{\ell ,i},i=1, \ldots, k_\ell,  \ell=1,2 \}$, that
is 
\be\label{ols} \hat \beta= (\hat \beta_0,\hat{\tilde \beta}_1,\hat{\tilde \beta}_2)=  \operatorname{arg\,min}\limits_{(b,,\tilde b_1,,\tilde b_2)\in B} \sum_{\ell=1}^2\sum_{i=1}^{k_\ell}\sum_{j=1}^{n_{\ell,i}}\big(Y_{\ell,i,j}-m_\ell(d_{\ell,i},(b_0,\tilde b_\ell))\big)^2.
\ee

\subsection{Testing equivalence of regression curves}\label{sec2.2}

Following \cite{liubrehaywynn2009} and  \cite{gsteiger2011}  we consider 
the regression curves $m_1$ and $m_2$  to be equivalent
if the maximum distance between the two curves is smaller than a given  pre-specified constant, say $ \epsilon >0  $,  that is, 
$$d_\infty(\beta_1,\beta_2) = \max_{d\in \mathcal{D}} |m_1(d,\beta_1)-m_2(d,\beta_2)| < \epsilon .
$$
In clinical trial practice $\varepsilon$ is often referred to as a relevance threshold in the sense that if $d_\infty(\beta_1,\beta_2)  < \epsilon $ the difference between the two curves is believed not to be clinically relevant.
In order to establish equivalence of the two curves $m_1$ and $m_2$ at a controlled type I error, we will develop a test for  the hypotheses
\begin{equation} \label{h0inf}
H_0: d_\infty (\beta_1, \beta_2) \geq \varepsilon \quad \mbox{versus} \quad H_1: d_\infty (\beta_1, \beta_2) < \varepsilon.
\end{equation}


In the following we extend the bootstrap approach from \cite{detmolvolbre2015} to test the hypotheses \eqref{h0inf} in the situation of common parameters.
Note that the test procedure proposed below could also be applied to alternative  measures of equivalence, such as the integrated deviation
$\int_{\cal D}|m_1(t,\beta_1)-m_2(t,\beta_2)| dt$.

\begin{algorithm} \label{alg1} {\rm (parametric bootstrap for testing equivalence under the assumption of common parameters)
		\begin{itemize}
			\item[(1)]
			Calculate the ordinary least-square (OLS) parameter estimate
			\eqref{ols}  assuming a common parameter $\beta_0$. The corresponding variance estimates are given by
			\be\label{var}
			\hat \sigma^2_\ell = \frac {1}{n_\ell} \sum^{k_\ell}_{i=1} \sum^{n_{\ell,i}}_{j=1} (Y_{\ell,i,j}- m_\ell (d_{\ell,i}, \hat \beta_\ell ))^2, \qquad \ell=1,2,
			\ee
			where $\hat \beta_\ell=(\hat\beta_0,\hat {\tilde \beta}_\ell),\ \ell=1,2$.
			Calculate the estimate 
			$$\hat d_\infty=d_\infty (\hat \beta_1, \hat \beta_2)= \max_{d\in \mathcal{D}} |m_1(d,\hat \beta_1)-m_2(d,\hat \beta_2)|  $$ 
			for the maximal deviation between the two  regression curves.
			\item[(2)] Define the  constrained estimates
			\begin{equation} \label{MLcons}
			{\hat{\hat{\beta}}_{\ell}}= \left\{
			\begin{array} {ccc}
			\hat \beta_\ell & \mbox{if} & \hat d_\infty \geq \varepsilon \\
			\bar \beta_\ell & \mbox{if} & \hat d_\infty < \varepsilon
			\end{array}  \right., \qquad \ell=1,2,
			\end{equation}
			where $\bar \beta_1, \bar \beta_2$ minimize the objective function in \eqref{ols} under the additional restriction 
			\begin{equation}\label{constr}
			d_\infty(\beta_1,\beta_2)=\max_{d\in \mathcal{D}} |m_1(d,\beta_1)-m_2(d,\beta_2)| = \epsilon.
			\end{equation}
			Define ${\hat{\hat{d}}_{\infty}} = d_\infty ({\hat{\hat{\beta}}_{1}}, {\hat{\hat{\beta}}_{2}})$ and note that ${\hat{\hat{d}}_{\infty}}\geq \epsilon$.
		\end{itemize}
		The next two steps describe the (parametric) bootstrap procedure.
		\begin{itemize}
			\item[(3)] Generate data
			\begin{equation}\label{bootdata}
			Y_{\ell,i,j}^*=m_\ell(d_{\ell,i},(\hat {\hat{\beta}}_0,\hat {\hat{\tilde\beta}}_\ell))+\eta_{\ell,i,j}^* ~,i=1,\ldots, n_{\ell,i},~\ell=1,2 ,
			\end{equation}
			with independent and normally distributed errors  $\eta_{\ell,i,j}^* \sim \mathcal{N}(0, \hat  \sigma_\ell^2)$.
			\item[(4)] Calculate  the OLS estimate  $\hat \beta^*$ as in Step (1) and the test statistic
			$$\hat d^*_\infty= \max_{d\in \mathcal{D}} |m_1(d,\hat\beta_1^*)-m_2(d,\hat\beta_2^*)|,$$
			where $\beta_\ell^*=(\beta_0^*,\tilde\beta_\ell^*),\ \ell=1,2$.  The  $\alpha-$quantile of the distribution of the distribution of 
			the statistic $\hat d^*_\infty$ is denoted by $q_{\alpha}^*$ and the null hypotheses in \eqref{h0inf} is rejected, whenever 
			\begin{equation} \label{testInf}
			\hat d_\infty < \hat{q}_{\alpha}^{*}.
			\end{equation}
		\end{itemize}
		
		In practice the $\hat{q}_{\alpha}^{*}$ can be calculated  repeating  steps (3) and (4), say $B$ times, in order to obtain replicates $\hat d^*_{\infty,1}, \dots, \hat d^*_{\infty,B}$ of $\hat d^*_\infty$.  
		An estimate of  $ \hat{q}_{\alpha}^{*}$  then is defined by  $\hat{q}_{\alpha}^{(B)} :=  \hat d_\infty^{*(\lfloor B \alpha \rfloor )}$, where $\hat d^{*(1)}_\infty \le \ldots  \le \hat d^{*(B)}_\infty$
		denotes the corresponding order statistic, and this estimate is used in \eqref{testInf}
	}
\end{algorithm}

The following theorem states that this algorithm yields a valid test procedure. The proof is left to the Appendix \ref{sec5}.

\begin{theorem}
	\label{thm7}
	The test defined by \eqref{testInf} is  a consistent, asymptotic $\alpha$-level test. That is
	\begin{equation}
	\lim_{n_1,n_2 \rightarrow\infty}\mathbb{P} \big(\hat d_\infty<\hat q_{\alpha }^{*}\big)=1, \label{consistence2}
	\end{equation}
	whenever $d_{\infty} < \varepsilon$,  and
	\begin{equation}\label{type1error}
	\limsup_{n_1,n_2\rightarrow\infty}\mathbb{P}\big(\hat d_\infty<\hat q_{\alpha }^{*}\big)\leq\alpha.
	\end{equation}
	if  $d_{\infty} \geq \varepsilon$. 
\end{theorem}

\begin{remark} 
	\label{remplac}
	{\rm 
	The results presented in this section remain correct in trials with a common placebo group, where 
	 $n_0$ observations   are taken at dose level $d_{1}=0$ (corresponding to
	 placebo), which are modelled by the random variables  $Y_{0,1}, \ldots  , Y_{0,n_{0}}$. 
	 For the sake of a simple presentation  we consider   location-scale type models, such that the common effect at the placebo can easily be 
	 modelled, but we note that more general models can be considered as well introducing additional constraints for the parameter.
	 
To be precise, we assume  that the  models in \eqref{mod1a} are given by 
		\begin{equation}\label{loc_scale}
m_\ell (d,\beta_\ell )=\beta_{0,1}+\tilde\beta_{\ell,1}\cdot m_{\ell}^0(d,\tilde\beta^0_\ell ),\ \ell=1,2,\  i=1,\ldots,k_\ell.
	\end{equation}
	where $m_{\ell}^0 (0 ,\tilde\beta^0_\ell )=0$   $(\ell=1,2)$, such that
	the condition $m_1 (0,\beta_\ell )=m_{2} (0,\beta_\ell )=\beta_{0,1}$ reflects the  fact that  there is only one placebo group (and as a consequence 
	a common placebo parameter).    Models of this type cover the most frequently used functional forms used in drug development and  several examples can be found in \cite{ting2006dose}. 
	Beside the location parameter $\beta_{0,1}$ there may be also other shared parameters, which we do not reflect in our notations for a better readability.
	 The $\ell$-th model is completely characterized by its parameter $\beta_\ell=(\beta_{0},\tilde{\beta_{\ell}})=(\beta_{0},\tilde\beta_{\ell,1},\tilde\beta^0_{\ell})$, $\ell=1,2$, and we obtain estimates of the model 
	 parameters by minimizing the sum of squares
	\begin{equation}\label{olc_plac}
	\hat \beta= (\hat \beta_0,\hat{\tilde \beta}_1,\hat{\tilde \beta}_2)=  \operatorname{arg\,min}_{b\in B} \sum_{j=1}^{n_{0}}\big(Y_{0,j}-b_{0,1}\big)^2+\sum_{\ell=1}^2\sum_{i=2}^{k_\ell}\sum_{j=1}^{n_{\ell,i}}\big(Y_{\ell,i,j}-(b_{0,1}+\tilde b_{\ell,1}\cdot m_{\ell}^0(d_{\ell ,i},\tilde b^0_\ell ))\big)^2.
	\end{equation}
	Theorem \ref{thm7}  remains valid in this situation and a proof can be found in the  Appendix  (see Section \ref{sec64}).
	}
\end{remark}

\subsection{Testing equivalence of model parameters}\label{sec2.3}

So far we assumed that the two regression models $m_1$ and $m_2$ share the common parameter $\beta_0$. In practice it may be necessary to assess whether this assumption is plausible using an appropriate equivalence test for the shared model parameters. To be more precise, we recall the definition  the parameters   $\beta_{\ell} $ in  model \eqref{mod1a}, i.e.
$$
\beta_\ell=(\beta_{\ell,1},\ldots,\beta_{\ell,p'},\ldots, \beta_{\ell,p_\ell})~,~~ \ell=1,2.
$$
and note that  assumption \eqref{common_parameters}  of $p^{\prime}$ common parameters  in the models $m_{1}$ and $m_{2}$ 
can be represented as  $(\beta_{1,1},\ldots,\beta_{1,p'})=(\beta_{2,1},\ldots,\beta_{2,p'})$ for $\ell =1,2$. In order to investigate if this assumption holds 
at least approximately we construct a test  for the hypotheses
\begin{equation} \label{hyp_pretest}
K_0: \max_{i=1,\ldots, p'}|\beta_{1,i}-\beta_{2,i}| \geq \delta
\quad \mbox{ versus } \quad
K_1: \max_{i=1,\ldots, p'}|\beta_{1,i}-\beta_{2,i}| < \delta,
\end{equation}
where $\delta$ denotes the equivalence margin.  
To be precise let $\hat \beta^{(\ell)}$ denote the least squares estimates  in model $m_{\ell}$ for the sample 
$\{ Y_{\ell ,i,j} : ~j=1,\ldots , n_{\ell ,i},i=1, \ldots, k_\ell \}$ ($ \ell=1,2$), and assume that for large sample considerations
the  sample sizes $n_\ell$  and $n_{\ell,i}$ converge to infinity such that
\be
\label{des}
\lim_{n_\ell \to \infty }  {n_{\ell,i} \over n_\ell}  &=& \zeta_{\ell,i} > 0 ~,~~i=1 , \ldots , k_{\ell}~,~\ell=1,2,
\\
\label{lambda}
\lim_{n_1,n_2\rightarrow \infty}\frac{n}{n_1} &=&  \lambda\in (1,\infty).
\ee
Under standard assumptions, which are listed in Section \ref{sec5}  it can be shown that  the least squares estimate 
$\hat \beta^{(\ell)}$ of the parameter $\beta_{\ell}$   in model $m_{\ell}$  
is approximately  normal distributed, that is 
\be\label{Sigma_common}
\sqrt{n_{\ell}}(\hat \beta^{(\ell)}-\beta_{\ell})\stackrel{\mathcal{D}}{\rightarrow}\mathcal{N}(0,\Sigma^{-1}_{\ell}) ~,~\ell=1,2,
\ee
where the symbol $\dn$ means  convergence in distribution and the matrix $\Sigma_{\ell}$ is defined by
\begin{equation}\label{sigmal}
\Sigma_\ell ={1 \over \sigma_\ell ^2} \sum_{i=1}^{k_\ell}\zeta_{\ell,i}\tfrac {\partial}{\partial b_\ell}  m_\ell (d_{\ell,i,}, b_\ell )\big|_{b_\ell=\beta_\ell}
\big(\tfrac {\partial}{\partial b_\ell}  m_\ell (d_{\ell,i,}, b_\ell )\big|_{b_\ell=\beta_\ell} \big)^T~,~~\ell=1,2.
\end{equation}
Here  and throughout this paper we assume that  the matrices $\Sigma_{1}$ and $\Sigma_{2}$ are non-singular. 
Consequently the difference $\sqrt{n} ( \hat \beta^{(1)}  - \hat \beta^{(2)})$
is also asymptotically normal distributed, and in particular
it follows for the first $p^{\prime}$ components of the difference that 
\begin{equation}\label{asym_pretest}
\sqrt{n}\big((\hat\beta_{1,1},\ldots,\hat\beta_{1,p'})-(\hat\beta_{2,1},\ldots,\hat\beta_{2,p'})-\big((\beta_{1,1},\ldots,\beta_{1,p'})-(\beta_{2,1},\ldots,\beta_{2,p'})\big)\big)
\stackrel{\mathcal{D}}{\rightarrow}\mathcal{N}(0,\Omega),
\end{equation}
where the matrix $\Omega$ is defined by 
\be
\label{omega}
\Omega:=\lambda\Lambda_1^{-1}+\tfrac{\lambda}{\lambda-1}\Lambda_2^{-1},
\ee
$\Lambda_\ell^{-1}=\big((\Sigma_\ell^{-1})_{ij})\big)_{i,j=1}^{p'}$ denotes the upper-left $p'\times p'$-block of the matrix $\Sigma_\ell^{-1}$ $(\ell=1,2)$
and $\lambda$ is defined in \eqref{lambda}.
Therefore we obtain the approximation 
$$
(\hat\beta_{1,1},\ldots,\hat\beta_{1,p'})-(\hat\beta_{2,1},\ldots,\hat\beta_{2,p'})\stackrel{\cal D}{\approx} \mathcal{N}((\beta_{1,1},\ldots,\beta_{1,p'})-(\beta_{2,1},\ldots,\beta_{2,p'}),\tfrac{1}{n}\Omega)~,
$$
where $\Omega$ is defined in \eqref{omega}.
We  can now  apply the test $(2.2)$ proposed in \cite{wang1999} by rejecting  the null hypothesis $K_0$  in \eqref{hyp_pretest}, whenever
\be\label{multivariate_equivalence}
|\hat\beta_{1,i}-\hat\beta_{2,i}|<\delta-t_{1-\alpha,n-2}\big(\tfrac{\hat\Omega_{ii}}{n(n-2)}\big)^{1/2}\text{ for all $i=1,\ldots,p'$},
\ee
where $t_{1-\alpha,n-2}$ denotes the $1-\alpha$ quantile of the $t$-distribution with $n-2$ degrees of freedom and $\hat\Omega_{ii}$ the $i$th diagonal element of the matrix $\hat\Omega$ which is an estimate for the (unknown) covariance matrix $\Omega$ (this is obtained by replacing the unknown parameters  $\beta_{\ell}$, $\sigma^{2}_{\ell}$ and weights $\zeta_{\ell,i}$ in \eqref{sigmal} by their 
corresponding
estimates and $n_{\ell,i}/n_{\ell}$, respectively).

\section{Finite sample properties} \label{sec3}
\def\theequation{3.\arabic{equation}}
\setcounter{equation}{0}
\setcounter{table}{0}

We now investigate the finite sample properties of the bootstrap test proposed
in Section \ref{sec2.2} in terms of power and size using numerical simulations.
The data is generated as follows:
\begin{enumerate}[leftmargin=1.6cm]
	\item[\textbf{(a)}] We choose the functional form of the models $m_{1}, m_{2}$ and specify their parameters $\beta_{1}, \beta_{2}$ (including a common parameter $\beta_0$), which determine the true underlying models. Further we choose variances
	$\sigma_\ell^2$ and the actual dose levels $d_{\ell,i}$, $\ell=1,2$.
	\item[\textbf{(b)}] For each dose $d_{\ell,i}$ we calculate $n_{\ell,i}$ values for the response given by $m_{\ell}(d_{\ell,i},(\beta_0,\tilde\beta_{\ell}))$. By generating residual errors $\eta_{\ell,i,j}\sim N(0,\sigma_\ell^{2})$ we obtain the final response data
	\begin{equation}
	Y_{\ell,i,j} = m_{\ell}(d_{\ell,i},(\beta_0,\tilde\beta_{\ell}))+\eta_{\ell,i,j},\qquad j=1, \ldots, n_{\ell,i},\ i=1, \ldots k_\ell,\ \ell=1,2.
	\label{algorithmus1}
	\end{equation}
\end{enumerate}
The simulation results below were obtained using $1'000$ simulation runs, where $B=500$ bootstrap replications were used
to calculate  quantiles of the bootstrap test.

In the following, we report the simulations results for power and size under three different scenarios.
We consider the four-parameter sigmoid Emax model
\begin{equation}\label{sigEmax}
m(d,\beta)=\beta_{1} +\frac{ \beta_{2}d^{\beta_{3}}}{\beta_{4}^{\beta_{3}}+d^{\beta_{3}}},
\end{equation}
which is frequently used in practice when modeling dose response relationships (see for example  \cite{gabr:wein:2007} or  \cite{thom:swee:soma:2014}). In model \eqref{sigEmax}
the parameter $\beta=(\beta_1,\beta_2,\beta_3,\beta_4)$ corresponds (in this order) to the placebo effect $E_0$, the maximum effect $E_{max}$, the Hill parameter $h$ determining the steepness of the dose-response curve and the dose $ED_{50}$ producing half of the maximum effect (\cite{macdougall2006}).
In what follows we add an index $\ell=0$ for a shared parameter or $\ell=1,2$ for the group under consideration.

\begin{figure}[h!]
	\begin{center}
		\begin{tabular}{cc}
			$(a)$ Scenario 1 & $(b)$ Scenario 2\\
			\includegraphics[width=0.5\textwidth]{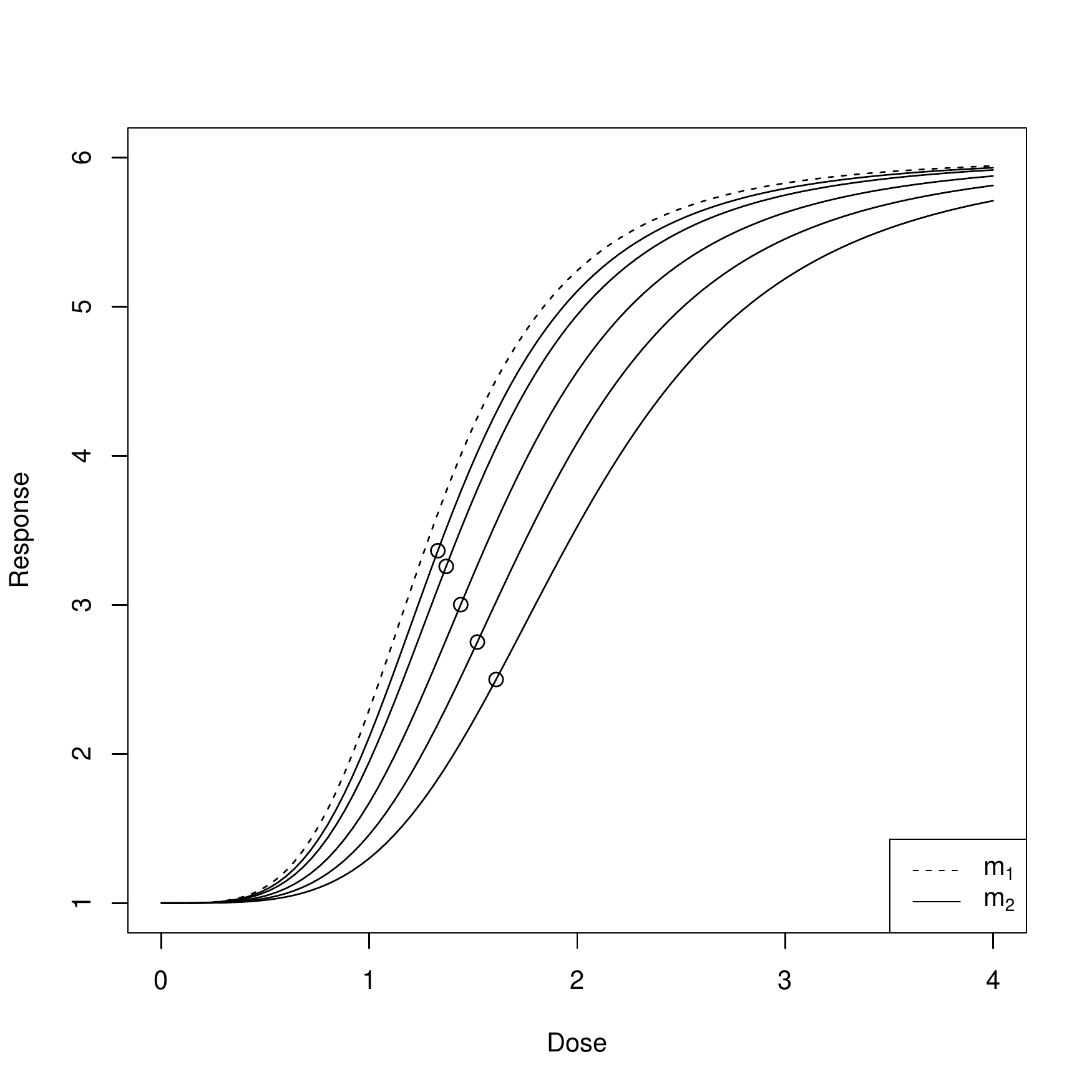} &
			\includegraphics[width=0.5\textwidth]{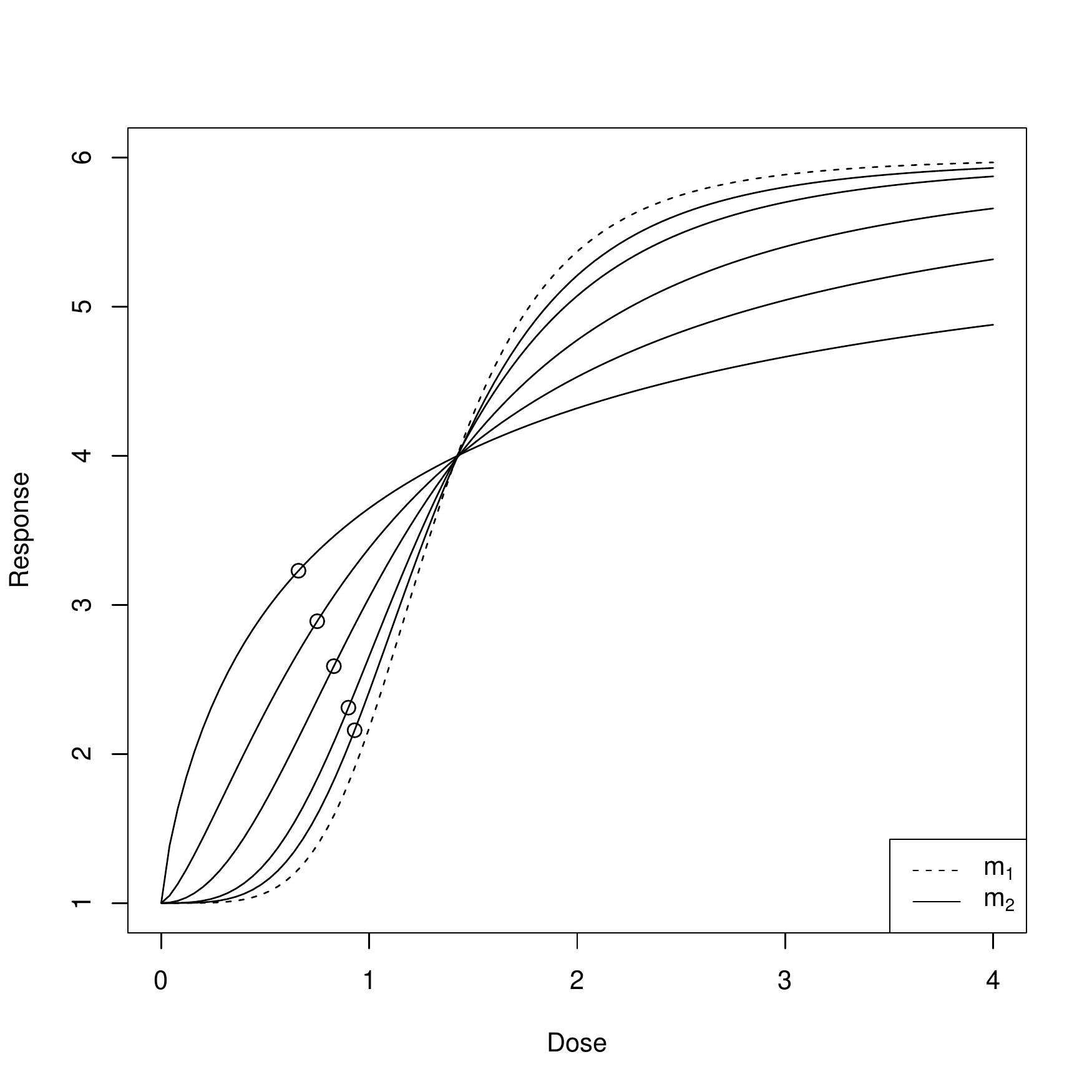}
		\end{tabular}
	\end{center}
	\caption{\small \it Graphical illustration of Scenarios~1 and~2.
		Open dots indicate the doses and corresponding responses where the maximum distance to the reference curve $m_1$ (dashed line) is attained.}
	\label{fig3}
\end{figure}

\textbf{Scenario 1:}
We assume the dose range $\mathcal{D}=[0,4]$ with identical dose levels $d_{\ell,i} = i-1$, $i=1,2,3,4,5$
for both regression models $\ell = 1, 2$.
For each configuration of $\sigma_\ell^2 = 1, 2, 3$
we use \eqref{algorithmus1} to simulate
$n_{\ell,i}=6,18,30$ observations at each dose level $d_{\ell,i}$, resulting in total sample sizes of $n_{\ell} = 30,90,150$, respectively.
We first compare the two sigmoid Emax models
\begin{equation}\label{ex3a}
m_1(d,\beta_1)=\beta_{0,1} +\frac{ \beta_{0,2}d^{\beta_{0,3}}}{\tilde\beta_{1,4}^{\beta_{0,3}}+d^{\beta_{0,3}}}  \qquad \mbox{and} \qquad m_2(d,\beta_2)=\beta_{0,1} +\frac{ \beta_{0,2}d^{\beta_{0,3}}}{\tilde\beta_{2,4}^{\beta_{0,3}}+d^{\beta_{0,3}}}.
\end{equation}
assuming the shared parameters $(\beta_{0,1},\beta_{0,2},\beta_{0,3})$.
The only difference between the two models is in the $ED_{50}$ parameters $\tilde\beta_{1,4}$ and $\tilde\beta_{2,4}$, which results in the need to estimate five parameters in total. We consider the reference sigmoid Emax model $m_1$ with the parameters 
$(\beta_{0,1}, \beta_{0,2}, \beta_{0,3}) = (1,5,4)$
and $\tilde \beta_{1,4}=1$. This reference model is compared to various specifications of the second model $m_2$ determined by $ \tilde \beta_{2,4}=1.99$, $1.77$, $1.59$, $1.43$, $1.37$, $1$ and common shared parameters $(\beta_{0,1}, \beta_{0,2}, \beta_{0,3})$. The values for $\tilde \beta_{2,4}$ were chosen
such that the maximum absolute distances $d_\infty=\max_{d\in{\cal D}}\left | m_{2}(\mathbf{\beta}_{2},d)-m_{1}(\mathbf{\beta}_{1},d)\right |$ are given by $2$, $1.5$, $1$, $0.5$, $0.25$, $0$ respectively. For $d_\infty>0$ these are attained at the dose levels $1.61,\ 1.52,\ 1.44,\ 1.37$ and $1.33$; see Figure \ref{fig3}$a$. For $d_\infty=0$,  that is $\tilde\beta_{1,4}=\tilde\beta_{2,4}$, the maximum distance is attained at every point in $\mathcal{D}$.

In Table  \ref{tab1a} we summarize the simulated rejection probabilities of
the bootstrap test  \eqref{testInf} under  the  null hypothesis \eqref{h0inf} with $d_\infty=2, 1.5, 1$ and $\epsilon=1$.
We conclude that the bootstrap test controls its level in all cases under consideration.
At the margin of the null hypothesis (i.e. $d_\infty=1$) the approximation of the level is very precise, even for sample sizes as small as $n_{\ell,i}=6$.

\begin{table}[h!]
	{\scriptsize
		\centering
		\begin{tabular}{||c|c|ccc||ccc||}
			\hline
			\multicolumn{1}{||c|}{} & \multicolumn{1}{|c|}{} & \multicolumn{3}{|c||}{$\alpha=0.05$} & \multicolumn{3}{|c||}{$\alpha=0.1$} \\\hline
			$n_\ell $  & $d_\infty$ & $\sigma^2=1$ & $\sigma^2=2$ & $\sigma^2=3$ &
			$\sigma^2=1$ & $\sigma^2=2$ & $\sigma^2=3$
			\\ \hline
			$30$  & 2 &  0.002 (0.000) & 0.010 (0.005) & 0.012 (0.009) & 0.002 (0.004)& 0.018 (0.013)& 0.031 (0.019) \\
			$30$  & 1.5 & 0.014 (0.013)& 0.035 (0.028) & 0.027 (0.043)& 0.026 (0.020) & 0.055 (0.049)& 0.052 (0.070)\\
			$30$  & 1    &  0.068 (0.058)& 0.054 (0.065)& 0.051 (0.060)& 0.106 (0.099) & 0.109 (0.114)& 0.115 (0.121)\\
			\hline
			$90$  & 2 & 0.000 (0.000)& 0.000 (0.000)& 0.001 (0.002)& 0.000 (0.000)& 0.000 (0.000)& 0.002 (0.002)\\
			$90$   & 1.5 & 0.004 (0.001)& 0.010 (0.006)& 0.013 (0.015)& 0.005 (0.005)& 0.021 (0.014)& 0.026 (0.020)\\
			$90$   &1    & 0.048 (0.071)& 0.053 (0.043)& 0.062 (0.059)& 0.104 (0.122)& 0.101 (0.097)& 0.129 (0.117)\\
			\hline
			$150$ & 2 & 0.000 (0.000)& 0.000 (0.000)& 0.000 (0.000)& 0.000 (0.000)& 0.000 (0.000)& 0.000 (0.004)\\
			$150$ & 1.5 & 0.000 (0.000)& 0.003 (0.000)& 0.002 (0.000)& 0.002 (0.000)& 0.011 (0.002)& 0.012 (0.009)\\
			$150$ &1    & 0.056 (0.061)& 0.040 (0.061)& 0.042 (0.063)& 0.103 (0.102)& 0.090 (0.102)& 0.096 (0.109) \\
			\hline
			\hline
		\end{tabular}
		\caption{\label{tab1a} \it Simulated Type I error   of the bootstrap test  \eqref{testInf}  for the equivalence of two
			sigmoid Emax models defined in Scenario~1 with $\epsilon=1$. The numbers in brackets show the simulated
			Type I error   when fixing the Hill parameter at $\beta_{0,3}=4$. }
	}
\end{table}

We also investigated the relative residual mean squared errors (RRMSE) of the parameters estimates. Table~\ref{tab1c} summarizes the simulation results only for $d_\infty=1$ (i.e. at the margin of the null hypothesis), as the results are similar for other choices of $d_\infty$. We conclude that the RRMSE for estimating the Hill parameter $\beta_{0,3}$ is (by far) the largest. This phenomenon has also been observed by \cite{mielke2016study}. We also observe that all estimation errors decrease with larger sample sizes and smaller variances. Table~\ref{tab1c} also summarizes the RRMSE when fixing the Hill parameter at $\beta_{0,3}=4$ (see the numbers in brackets). In this case, four parameters need to be estimated in total and the estimation errors become slightly smaller. We also repeated the Type I error rate simulations when fixing the Hill parameter at $\beta_{0,3}=4$. The the results are reported in Table~\ref{tab1a} (numbers in brackets) and we conclude that the size is well controlled within the simulation error.

\begin{table}[h!]
	{\scriptsize
		\centering
		\begin{tabular}{||c|c|ccccc||}
			\hline
			$n_\ell $ & $\sigma^2$ & $\beta_{0,1}$ & $\beta_{0,2}$ & $\beta_{0,3}$ & $\tilde\beta_{1,4}$ & $\tilde\beta_{2,4}$	\\ \hline
			$30$ & 1 &0.288 (0.263) & 0.091 (0.062)  &0.493 & 0.125 (0.108) & 0.114 (0.106) \\
			$30$ & 2 &0.389 (0.358)& 0.118 (0.090)& 0.739 & 0.174 (0.157)& 0.172 (0.148) \\
			$30$ & 3 &0.460 (0.427)& 0.134 (0.105) & 0.841 & 0.228 (0.198)& 0.211 (0.185) \\
			\hline
			$90$ & 1 & 0.166 (0.152) & 0.054 (0.036) & 0.184 & 0.067 (0.064)& 0.063 (0.061)\\
			$90$ & 2 &	0.237 (0.219) & 0.076 (0.054)& 0.355 & 0.107 (0.086) & 0.091 (0.086)\\
			$90$ & 3&	0.280 (0.261)& 0.091 (0.062)&0.450 & 0.130 (0.112)& 0.113 (0.105) \\
			\hline
			$150$ & 1 &0.123 (0.115)& 0.040 (0.029) & 0.129 & 0.050 (0.057)& 0.049 (0.046) \\
			$150$ &	2 & 0.171 (0.170)  &0.057 (0.041) & 0.241 & 0.073 (0.072) & 0.068 (0.063) \\
			$150$ &3&	0.219 (0.204)& 0.070 (0.050)& 0.322 & 0.095 (0.086) & 0.091  (0.084)\\ 	
			\hline
			\hline
		\end{tabular}
		\caption{\label{tab1c} \it RRMSE of the parameters obtained in the model estimation step of the bootstrap test  \eqref{testInf}
			for the equivalence of two sigmoid Emax models defined in Scenario~1 with $d_\infty=1$. The numbers in brackets show the values for the RRMSE when fixing the Hill parameter at $\beta_{0,3}=4$.}
	}
\end{table}

In Table \ref{tab1b} we summarize the power of the bootstrap when generating the data under the alternative $d_\infty=0.5, 0.25, 0$ and $\epsilon=1$.
As expected, the power increases with larger sample sizes and smaller variances and is reasonably high across all configurations. Fixing the Hill parameter significantly improves the power which can be explained by the difficulty of estimating this parameter precisely, as discussed above.

\begin{table}[h!]
	{\scriptsize
		\centering
		\begin{tabular}{||c|c|ccc||ccc||}
			\hline
			\multicolumn{1}{||c|}{} & \multicolumn{1}{|c|}{} & \multicolumn{3}{|c||}{$\alpha=0.05$} & \multicolumn{3}{|c||}{$\alpha=0.1$} \\\hline
			$n_\ell $ & $d_\infty$ & $\sigma^2=1$ & $\sigma^2=2$ & $\sigma^2=3$ &
			$\sigma^2=1$ & $\sigma^2=2$ & $\sigma^2=3$
			\\ \hline
			$30$   & 0.5 & 0.137 (0.154) & 0.075 (0.078) & 0.070 (0.073) & 0.238 (0.266) & 0.172 (0.155) & 0.137 (0.145)\\
			
			$30$  & 0.25 & 0.208 (0.190)& 0.102 (0.101)& 0.081 (0.088)& 0.344 (0.349) & 0.196 (0.188) & 0.152 (0.170) \\
			$30$ & 0     & 0.181 (0.203) & 0.105 (0.105)& 0.086 (0.092) & 0.333 (0.361) & 0.196 (0.213)& 0.154 (0.154) \\
			\hline
			$90$  & 0.5 & 0.341 (0.424)& 0.180 (0.230)& 0.132 (0.153)& 0.505 (0.581)& 0.311 (0.357)& 0.246 (0.279)\\
			$90$  & 0.25 & 0.550 (0.675)& 0.249 (0.315)& 0.166 (0.190)& 0.733 (0.802)& 0.428 (0.484)& 0.305 (0.348)\\
			$90$  & 0 & 0.664 (0.783)& 0.286 (0.353)& 0.191 (0.188)& 0.822 (0.884)& 0.463 (0.562)& 0.338 (0.367) \\
			\hline
			$150$  & 0.5 & 0.481 (0.593) & 0.297 (0.359) & 0.207 (0.273)& 0.635 (0.729)& 0.460 (0.502)& 0.355 (0.406) \\
			$150$ & 0.25 & 0.826 (0.868)& 0.448 (0.569)& 0.280 (0.357)& 0.902 (0.933) & 0.635 (0.719)& 0.477 (0.545)\\
			$150$ & 0 & 0.917 (0.961)& 0.559 (0.665)& 0.342 (0.415)& 0.966 (0.989)& 0.740 (0.812)& 0.520 (0.596)\\
			
			\hline
			\hline
		\end{tabular}
		\caption{\label{tab1b} \it Simulated power of the bootstrap test  \eqref{testInf}
			for the equivalence of two sigmoid Emax models defined in Scenario~1 with $\epsilon=1$. The numbers in brackets show the simulated
			power when fixing the Hill parameter at $\beta_{0,3}=4$.  }
	}
\end{table}

\textbf{Scenario 2:}
We maintain the basic settings from Scenario $1$. We consider again two sigmoid Emax models
\begin{equation}\label{ex3b}
m_1(d,\beta_1)=\beta_{0,1} +\frac{ \beta_{0,2}d^{\tilde\beta_{1,3}}}{\tilde\beta_{1,4}^{\tilde\beta_{1,3}}+d^{\tilde\beta_{1,3}}}  \qquad \mbox{and} \qquad m_2(d,\beta_2)=\beta_{0,1} +\frac{\beta_{0,2}d^{\tilde\beta_{2,3}}}{\tilde\beta_{2,4}^{\tilde\beta_{2,3}}+d^{\tilde\beta_{2,3}}}.
\end{equation}
but assume now that both, the placebo response $\beta_{0,1}$ and the maximum treatment effect $\beta_{0,2}$, are the same.
For the reference model we chose $(\beta_{0,1},\beta_{0,2})=(1,5)$, $\tilde\beta_{1,3}=4.5$ and $\tilde\beta_{1,4}=1.3$.
We investigated the maximum distances
$d_\infty =2,1.5, 1, 0.5, 0.25, 0$, which resulted in the following parameter configurations for the second model:
{\small \begin{align}\label{scenario2}
	(\tilde \beta_{2,3},\tilde \beta_{2,4})=(0.81,0.86),\ (\tilde \beta_{2,3},\tilde \beta_{2,4})=(1.4,1.07),\ (\tilde \beta_{2,3},\tilde \beta_{2,4})=(2.15,1.18),\nonumber\\  (\tilde \beta_{2,3},\tilde \beta_{2,4})=(3.15,1.25),\ (\tilde \beta_{2,3},\tilde \beta_{2,4})=(3.75,1.28),\ (\tilde \beta_{2,3},\tilde \beta_{2,4})=(\tilde \beta_{1,3},\tilde \beta_{1,4})=(4.5,1.3).
	\end{align}}
The maximum distances  between the two curves  are now attained at the dose levels $0.66,\ 0.75,\ 0.83,\ 0.9$ and $0.93$;
see Figure \ref{fig3}$b$.

In Table~\ref{tab2a} we summarize the simulated rejection probabilities under the null hypothesis \eqref{h0inf} for $d_\infty =2,1.5, 1$ and $\epsilon=1$. We conclude again that the bootstrap test controls the designated significance level in all cases under consideration. Especially at the margin $d_\infty =1$ the simulated Type I error rates are close to the nominal level $\alpha$. These observations apply regardless of whether the Hill parameter is estimated or fixed at the true underlying values given in \eqref{scenario2}.

\begin{table}[h!]
	{\scriptsize
		\centering
		\begin{tabular}{||c|c|ccc||ccc||}
			\hline
			\multicolumn{1}{||c|}{} & \multicolumn{1}{|c|}{} & \multicolumn{3}{|c||}{$\alpha=0.05$} & \multicolumn{3}{|c||}{$\alpha=0.1$} \\\hline
			$n_\ell $  & $d_\infty$ & $\sigma^2=1$ & $\sigma^2=2$ & $\sigma^2=3$ &
			$\sigma^2=1$ & $\sigma^2=2$ & $\sigma^2=3$
			\\ \hline
			$30$  & 2    &  0.002 (0.000) & 0.001 (0.000) & 0.001 (0.000) & 0.000 (0.005)& 0.003 (0.007)& 0.006 (0.011) \\
			$30$  & 1.5  &  0.000 (0.001)& 0.003 (0.000) & 0.008 (0.001)& 0.007 (0.033) & 0.002 (0.030)& 0.016 (0.054)\\
			$30$  & 1    &  0.036 (0.035)& 0.044 (0.040)& 0.053 (0.039)& 0.083 (0.099) & 0.110 (0.116)& 0.113 (0.112)\\
			\hline
			$90$     & 2 & 0.000 (0.000)& 0.000 (0.000)& 0.000 (0.000)& 0.000 (0.000)& 0.000 (0.002)& 0.004 (0.002)\\
			$90$   & 1.5 & 0.000 (0.000)& 0.000 (0.000)& 0.004 (0.000)& 0.000 (0.007)& 0.004 (0.021)& 0.012 (0.022)\\
			$90$   &1    & 0.052 (0.057)& 0.028 (0.040)& 0.016 (0.036)& 0.104 (0.113)& 0.068 (0.117)& 0.056 (0.105)\\
			\hline
			$150$   & 2 & 0.000 (0.000)& 0.000 (0.000)& 0.000 (0.000)& 0.000 (0.000)& 0.000 (0.000)& 0.000 (0.000)\\
			$150$ & 1.5 & 0.000 (0.000)& 0.000 (0.000)& 0.004 (0.000)& 0.000 (0.000)& 0.000 (0.004)& 0.004 (0.004)\\
			$150$ &1    & 0.056 (0.032)& 0.036 (0.036)& 0.040 (0.036)& 0.100 (0.088)& 0.080 (0.100)& 0.076 (0.088) \\
			\hline
			\hline
		\end{tabular}
		\caption{\label{tab2a} \it Simulated Type I error   of the bootstrap test  \eqref{testInf}  for the equivalence of two
			sigmoid Emax models defined in Scenario~2 with $\epsilon=1$. The numbers in brackets show the simulated
			Type I error   when fixing the Hill parameters at their true underlying values. }
	}
\end{table}

In Table~\ref{tab2b} we summarize the simulated power of the bootstrap test under the alternative $d_\infty=0.5, 0.25, 0$ and $\epsilon=1$.
As expected, the power decreases for increasing values of $d_\infty$ and for higher variances or smaller sample sizes.
{One noticeable exception occurs at $d_\infty=0$, where in some cases the power is smaller than for $d_\infty=0.25$. 
	This effect can be explained theoretically when considering the proofs for the bootstrap test. In case of $d_\infty=0$ the set $\cal E=\cal E^+ \cup \cal E^-$ containing all points where the maximum distance between the two curves is attained (see Appendix \ref{sec63}) consists of the entire dose range $\cal D$. Therefore, the asymptotic distribution of the test statistic is not Gaussian but a maximum of Gaussian processes. This complex structure of the asymptotic distribution has an impact on the bootstrap procedure and explains the
	decrease in  power  for  $d_\infty=0$. This phenomenon can also be observed, although to a lesser degree, in Scenario $1$.}
Finally, we observe higher power values when fixing the Hill parameter compared to the situation where it has to be estimated.

\begin{table}[h!]
	{\scriptsize
		\centering
		\begin{tabular}{||c|c|ccc||ccc||}
			\hline
			\multicolumn{1}{||c|}{} & \multicolumn{1}{|c|}{} & \multicolumn{3}{|c||}{$\alpha=0.05$} & \multicolumn{3}{|c||}{$\alpha=0.1$} \\\hline
			$n_\ell $ & $d_\infty$ & $\sigma^2=1$ & $\sigma^2=2$ & $\sigma^2=3$ &
			$\sigma^2=1$ & $\sigma^2=2$ & $\sigma^2=3$
			\\ \hline
			$30$   & 0.5   & 0.130 (0.147) & 0.096 (0.075) & 0.094 (0.072) & 0.231 (0.245) & 0.191 (0.133) & 0.174 (0.125)\\
			
			$30$  & 0.25   & 0.156 (0.194)& 0.102 (0.089)& 0.087 (0.085)& 0.275 (0.319) & 0.197 (0.178) & 0.157 (0.147) \\
			$30$ & 0       & 0.155 (0.164) & 0.108 (0.076)& 0.087 (0.059) & 0.310 (0.316) & 0.197 (0.166)& 0.189 (0.137) \\
			\hline
			$90$  & 0.5   & 0.312 (0.542)& 0.124 (0.225)& 0.140 (0.133)& 0.528 (0.697)& 0.240 (0.384)& 0.260 (0.240)\\
			$90$  & 0.25  & 0.384 (0.689)& 0.224 (0.289)& 0.164 (0.173)& 0.560 (0.841)& 0.396 (0.484)& 0.292 (0.313)\\
			$90$  & 0     & 0.448 (0.663)& 0.192 (0.259)& 0.148 (0.163)& 0.616 (0.807)& 0.372 (0.455)& 0.240 (0.309) \\
			\hline
			$150$  & 0.5 & 0.528 (0.780) & 0.220 (0.392) & 0.160 (0.228)& 0.688 (0.896)& 0.404 (0.632)& 0.256 (0.400) \\
			$150$ & 0.25 & 0.724 (0.904)& 0.320 (0.544)& 0.260 (0.304)& 0.824 (0.936) & 0.540 (0.740)& 0.408 (0.516)\\
			$150$ & 0    & 0.644 (0.920)& 0.308 (0.580)& 0.224 (0.248)& 0.800 (0.956)& 0.532 (0.728)& 0.404 (0.484)\\
			
			\hline
			\hline
		\end{tabular}
		\caption{\label{tab2b} \it Simulated power of the bootstrap test  \eqref{testInf}
			for the equivalence of two sigmoid Emax models defined in Scenario~2 with $\epsilon=1$. The numbers in brackets show the simulated
			power when fixing the Hill parameters at their true underlying values.  }
	}
\end{table}

\begin{figure}[h!]
	\begin{center}
		\begin{tabular}{cc}
			$(a)$  & $(b)$ \\
			\includegraphics[width=0.49\textwidth]{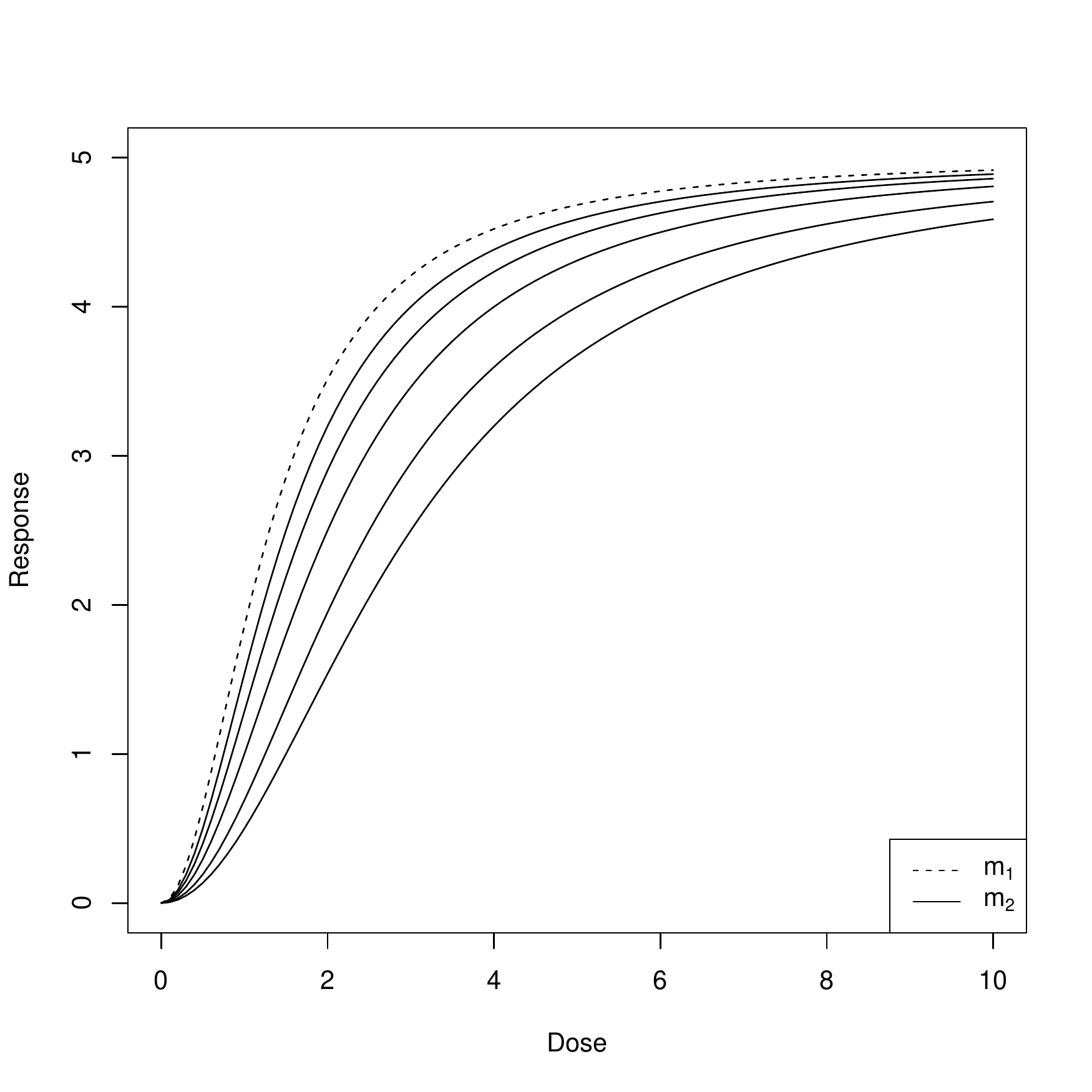} &
			\includegraphics[width=0.49\textwidth]{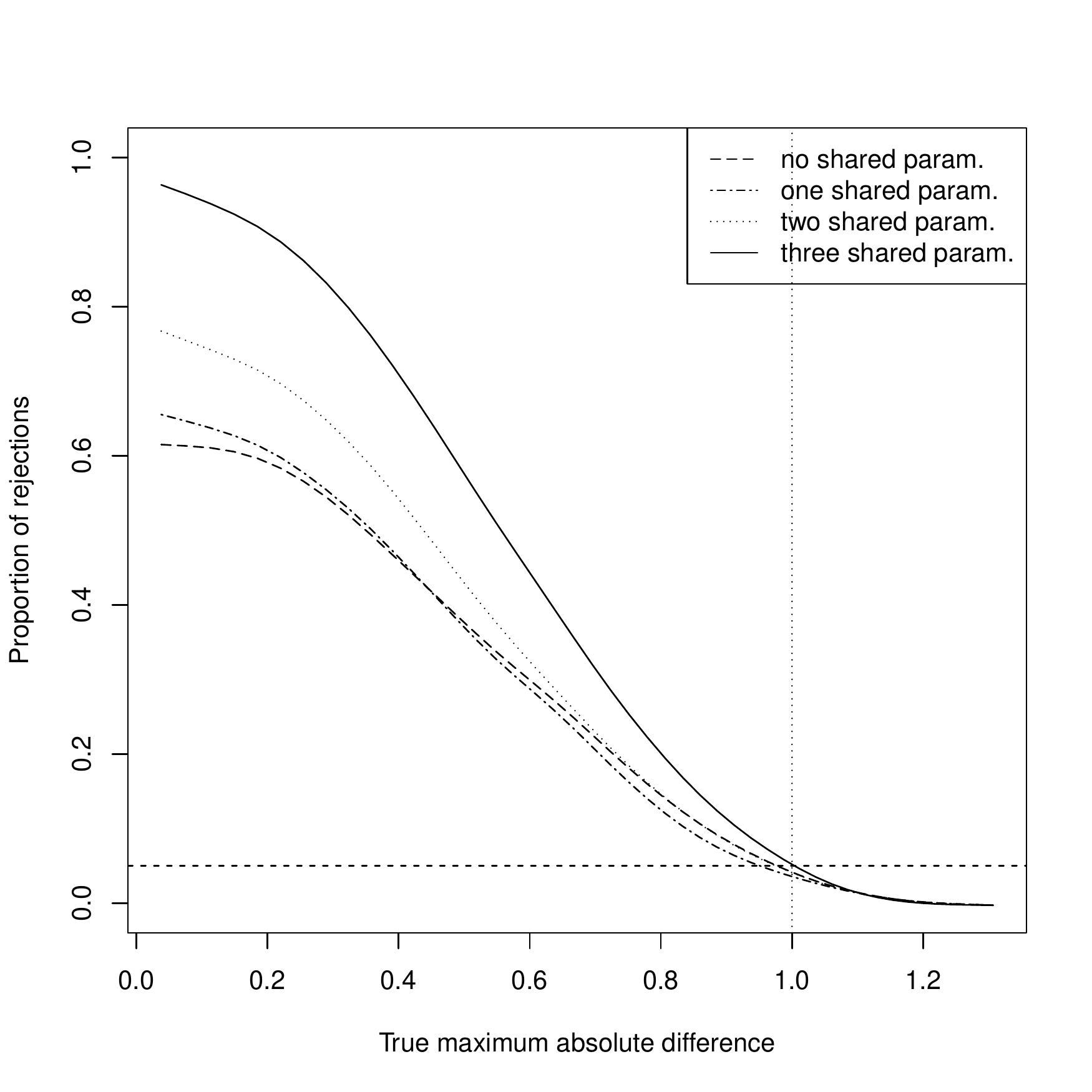}
		\end{tabular}
	\end{center}
	\caption{\small{\it  $(a)$ Graphical illustration of the regression functions  $m_1$ and $m_2$ with $\kappa=1.5,1.7,2,2.5,3$ in Scenario 3. $(b)$ Proportion of rejections in dependence of the true maximum absolute difference $d_\infty$ for four different tests assuming one, two, three and no shared parameters for $\epsilon=1$ (vertical dotted line) and $\alpha=0.05$ (horizontal dashed line).}
	}
	\label{fig4}
\end{figure}

\textbf{Scenario 3:}
We now investigate the operating characteristics of the bootstrap test assuming three, two, one and no shared parameters. We set $\epsilon=1$, $\alpha=0.05$ and compare again two sigmoid Emax models.
The true placebo response is chosen as $\beta_{1,1}=\beta_{2,1}=0$. The reference model $m_1$ is specified by $(\beta_{1,2},\beta_{1,3},\beta_{1,4})=(5,2,1.3)$. The second model is specified by $(\beta_{2,2}, \beta_{2,3},\beta_{1,4})=(5,2,\kappa)$, where $\kappa \in (1.3,3]$ is chosen such that the maximum distances with respect to $m_1$ lie between $0$ and $2$; see the resulting curves plotted in Figure \ref{fig4}$(a)$ for $\kappa=1.5,1.7,2,2.5,3$. Consequently the parameters specifying the two models only differ in $ED_{50}$ parameters $\tilde\beta_{1,4}$ and $\tilde\beta_{2,4}$ and the maximum distance is determined by the choice of $\kappa$. As all other parameters are the same for the two models, we can compare the bootstrap test assuming three, two, one and no shared parameter. Note that we do not consider the case of identical models (i.e. $\kappa=1.3$) because of the discontinuity of power at
$d_\infty=0$ described under Scenario 2.
The dose range is given by $\mathcal{D}=[0,10]$ with $5$ different dose levels $d_{\ell,1} =0$, $d_{\ell,2} =1$, $d_{\ell,3} =2$,
$d_{\ell,4} =5$, $d_{\ell,5} =10, \ell=1,2$.
We create $n_{\ell,i}=35$ observations at each dose level for each group according to \eqref{algorithmus1}, which results in a total sample size of $n=n_1+n_2=350$. 
Finally, we choose $\sigma_\ell^2=2$, $\ell=1,2$.  

In Figure \ref{fig4}$(b)$ we plot the proportion of rejections in dependence of the true maximum absolute difference $d_\infty \in (0,2]$. Under the null hypothesis  $d_\infty \geq 1$ all four tests control their level, as the proportion of rejections is smaller than or equal to $\alpha=0.05$ within simulation errors. Looking at the region $d_\infty <1$, we observe that the test assuming three shared parameters has the highest power among all four tests, followed by the test assuming two shared parameters. The difference between the tests assuming one and no shared parameter is rather small.
Concluding, the more parameters can be assumed to be common for the two regression curves the higher is the power of the test.
Note, however, that strictly speaking the hypotheses \eqref{h0inf} are different when assuming three, two, one and no shared parameters and that the perceived power gain when assuming more shared parameters comes at the cost of making additional assumptions that need to be verified in practice, as
illustrated with the clinical trial example in Section~\ref{sec4}.

\section{Clinical trial example} \label{sec4}
\def\theequation{4.\arabic{equation}}
\setcounter{equation}{0}

We now illustrate the proposed method with a multi-regional clinical trial example. The objective of this trial is to evaluate the dose response relationships in Caucasian and Japanese patients and assess their similarity. Based on data from previous clinical trials investigating a drug with a similar mode of action, it is reasonable to assume a similar response to placebo and a common maximum treatment effect in both populations, with the main difference expected to be in a different onset of treatment effect. Using the sigmoid Emax model \eqref{sigEmax}, these consideration thus lead to different $ED_{50}$ and Hill parameters for the two dose response curves. Because the trial is still at its design stage, we simulate data based on the trial assumptions. To maintain confidentiality, we scale the actual doses to lie within the [0, 15] interval. These limitations do not change the utility of the calculations below.

We assume $60$ Japanese and $240$ Caucasian patients, resulting in $300$ patients overall. Patients from both populations are randomized to receive either placebo (dose level $0$) or one of three active dose levels, namely $1,\ 3,\ 15$ for the Japanese and $0.5,\ 9$ and $15$ for the Caucasian patients. Assuming equal allocation of patients within each population, we thus have 75, 60, 15, 15, 60, and 75 patients randomized to the dose levels $0,\ 0.5, 1, 3, \ 9$ and $15$, respectively. The response variable is assumed to be normally distributed and larger values indicate a better outcome. Pharmacological and clinical considerations suggest the use of the (three-parameter) Emax model with the Hill parameter fixed at 1. Later on we relax this assumption as part of a sensitivity analysis. The \texttt{R} code for this example and all other calculations in this paper is available from the authors upon request.

\begin{figure}[h!]
	\begin{center}
		\includegraphics[width=0.6\textwidth]{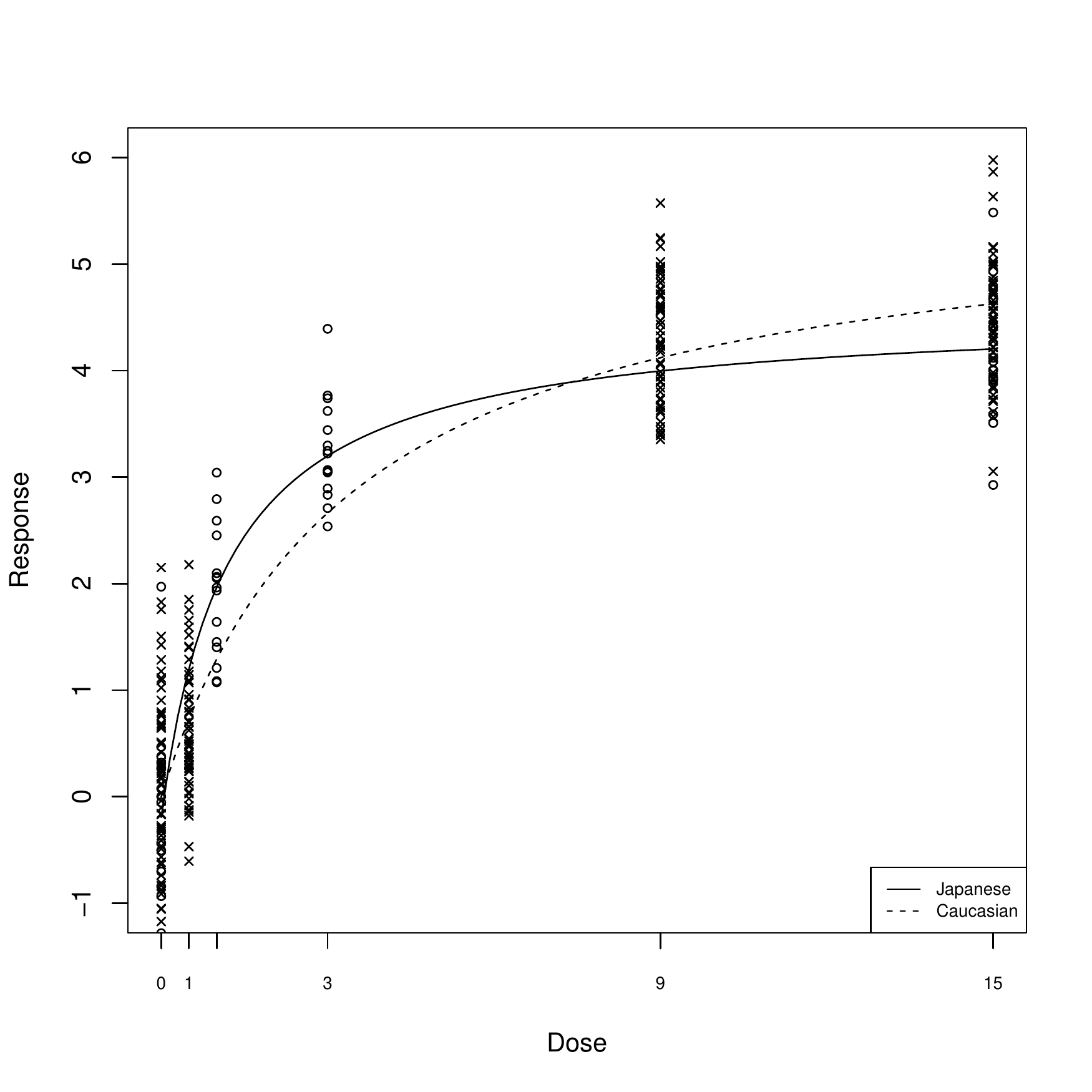}
	\end{center}
	\caption{\small \it 
		Fitted Emax model $m_1$ ($m_2$) for the Japanese (Caucasians) patients given by the solid (dashed) line with observations marked by ``o'' (``x'').}
	\label{fig2}
\end{figure}

In Figure~\ref{fig2} we display the fitted dose response models $m_1(d,\hat{\beta}_{1})$ and $m_2(d,\hat{\beta}_{2})$ for the Japanese and Caucasian patients, respectively, together with the individual observations, where $d \in [0, 15]$ and the $y$-axis is truncated to $[-1,6]$ for better readability. The parameter estimates from the two separate model fits are given by $\hat\beta_1=(-0.195,  4.751, 11.991)$ and $\hat\beta_2=(-0.002,  5.676, 33.887)$. The observed differences for the placebo response and the maximum treatment effect are given by $|\hat\beta_{1,1}-\hat\beta_{2,1}|=0.193$ and $|\hat\beta_{1,2}-\hat\beta_{2,2}|=0.925$, respectively, and thus relatively small, as it also transpires from the plots in Figure~\ref{fig2}. To corroborate this empirical observation, we formally test whether the assumption of shared parameters is plausible by applying the equivalence test described in Section~\ref{sec2.3} on the data set under consideration. We choose the threshold $\delta = 1.5$ and therefore test the null hypothesis $K_0: \max_{i=1,2}\left| \beta_{1,i}-\beta_{2,i}\right| \geq 1.5$ against the alternative $K_1:\max_{i=1,2}\left| \beta_{1,i}-\beta_{2,i}\right| < 1.5.$ Applying the test \eqref{multivariate_equivalence} for $\alpha=0.05$, we obtain $\hat\Omega_{11} =3127.91$ and $\hat\Omega_{22} = 10748.27$ and therefore $\delta-t_{1-\alpha,n-2}\big(\tfrac{\hat\Omega_{11}}{n(n-2)}\big)^{1/2} = 1.191$ and $\delta-t_{1-\alpha,n-2}\big(\tfrac{\hat\Omega_{22}}{n(n-2)}\big)^{1/2} = 0.928$, respectively. We can thus reject $K_0$ at the relatively stringent 5\% level and conclude equivalence of the two parameters, which justifies using the bootstrap test \eqref{alg1} with shared parameters.


We now evaluate the similarity of the dose response curves for the Japanese and Caucasian patients, assuming the same placebo and maximum treatment effect. In order to compute the non-linear least squares estimates in model  \eqref{mod1a} with \eqref{common_parameters} we formulate the objective function of the minimization step as
$$
\sum_{i=1}^{4}\sum_{j=1}^{60}\Big (Y_{1,i,j}-\Big (\beta_{0,1}  +\frac{ \beta_{0,2}d_{1,i,j}}{\tilde\beta_{1,3}+d_{1,i,j}}\Big )\Big )^2
+\sum_{i=1}^{4}\sum_{j=1}^{240}\Big (Y_{2,i,j}-\Big (\beta_{0,1} +\frac{ \beta_{0,2}d_{2,i,j}}{\tilde\beta_{2,3}+d_{2,i,j}}\Big )\Big )^2.$$
Here, $\beta_{0,1}$ denotes the (shared) placebo effect, $\beta_{0,2}$ the (shared) maximum treatment effect $E_{max}$, and $\tilde\beta_{1,3}$ and $\tilde\beta_{2,3}$ the $ED_{50}$ parameters of the two models.
Using the \texttt{auglag} function from the \texttt{alabama} package \cite{varadhan2014} to solve the above optimization problem, we obtain the parameter estimates
$\hat\beta_{0,1}=-0.064\ (0.074)$, $\hat\beta_{0,2}=5.366\ (0.137)$, $\hat{\tilde\beta}_{1,3}= 19.400\ (2.634)$ and $\hat{\tilde\beta}_{2,3}=25.681\ (3.256)$. In brackets we report the associated standard errors, which have to be calculated manually based on \eqref{convergence_common}. The estimates for the population variances  are $\hat\sigma_1^2=0.508$ and $\hat\sigma_2^2=0.455$. The observed maximum difference between both curves over the investigated dose range $[0,15]$ is
$\hat d_\infty=0.376 $, attained at dose $2.23$.
We apply the bootstrap test \eqref{testInf} using $B=1'000$ bootstrap replications. Setting $\varepsilon=0.7$ for the equivalence margin in \eqref{h0inf}, we
obtain the quantile
$q_{0.05} = 0.438$ for $\alpha=0.05$. Thus, we reject the null hypothesis \eqref{h0inf} at the 5\% significance level and conclude that the dose response curves for the Japanese and Caucasian populations are similar, under the shared parameter assumption.
Alternatively, we can calculate the $p$-value
$\frac{1}{B} \sum_{i=1}^B I (d_\infty^{*(i)}\leq \hat d_\infty)=0.023$ for the bootstrap test and obtain the same test decision at level $\alpha=0.05$.
For illustration purposes we also apply the bootstrap test \eqref{testInf} but without shared parameters (yet under the assumption of a fixed Hill parameter). Accordingly, we obtain a considerably larger $p$-value of $0.458$, which supports our findings from Scenario $3$ in Section \ref{sec3} about the loss in power when no shared parameters are assumed. In this case the observed maximum distance is $\hat d_\infty=0.706$, attained at dose $1.42$, and the quantile of the bootstrap distribution is $q_{0.05} = 0.449$.

Finally, we perform a sensitivity analysis to investigate the assumption of the Hill parameter being equal to 1. As part of this analysis we repeat the model fit and the bootstrap test using the sigmoid Emax model \eqref{sigEmax} where the Hill parameter is now part of the estimation.
The parameter estimates (standard errors in brackets) are
$\hat\beta_{0,1}=0.037\ (0.082)$, $\hat\beta_{0,2}=4.544\ (0.218)$, $\hat{\tilde\beta}_{1,3}= 1.05\ (0.229)$,  $\hat{\tilde\beta}_{1,4}= 13.542\ (2.095)$, $\hat{\tilde\beta}_{2,3}=1.650\ (0.331)$ and $\hat{\tilde\beta}_{2,4}=16.558\ (4.521)$. Now, the maximum distance between the curves is
$\hat d_\infty=0.640 $, attained at dose $0.6$.
It turns out that the standard errors of the estimates are slightly higher which is in line with the results shown in the simulation studies in Section \ref{sec3}.
Performing again the bootstrap test with two shared parameters
results the quantile $q_{0.05} = 0.429$ and the $p$-value $0.285$. Consequently, we cannot reject the null hypothesis in this case.
In conclusion, fixing the Hill parameter to 1 and assuming both the placebo effect and the maximum treatment effect to be the same in both populations clearly results in the most powerful procedure. We can demonstrate equivalence at the significance level of $\alpha=0.05$, whereas in case of estimating both models separately (i.e. no shared parameters) or including the Hill parameter in the estimation we obtain considerably larger $p$-values.

%

\section{Conclusion}
In this paper we developed a new test for the  equivalence  of two regression curves when it is reasonable to assume that some model parameters are the same.
Our approach is based on an estimate of the maximum deviation between the two curves, where critical values are obtained by a novel constraint bootstrap procedure.
We demonstrated that the new test controls its level properly and is consistent. 

We investigated the finite sample properties of the proposed procedure using extensive simulations and observed that the Type I error rate is controlled in all scenarios under consideration, even for sample sizes as small as $6$ patients per dose level. Further, we concluded that the test reaches a reasonable power that increases with larger sample sizes. In particular, we 
demonstrated that the power of tests for the equivalence of curves can be improved substantially by using the additional information 
of common  parameters in the two regression curves.
This effect could also be observed in the clinical trial example, which showed the power advantage of the bootstrap test \eqref{testInf} if the underlying assumptions are well justified. Relaxing those assumptions may lead to more robust conclusions, but only at a cost of a loss in power.

An interesting extension of the proposed methodology arises from the need to include covariates in clinical trial practice. Covariates can be continuous (e.g. age or body mass index), categorical (e.g. disease status or race), or binary (e.g. gender or smoking yes/no), possibly changing over time. These cases may have to be treated differently and we leave this problem for future research. Another area of research could be the assessment of similarity in two nested populations, thus relaxing the assumption of independence between the observations. In our multi-regional clinical trial example we compared the Japanese with Caucasian patients. It will be interesting and relevant to clinical trials to explore the development of the proposed methods when comparing the Japanese with an overall population that includes Japanese and Caucasian patients. Again, we leave this topic for future research.

\bigskip 
\textbf{Acknowledgments}
This work has been supported in part by the
Collaborative Research Center ``Statistical modeling of nonlinear
dynamic processes'' (SFB 823, Teilprojekt T1) of the German Research Foundation (DFG) 
Parts of this manuscript were written while Frank Bretz was on a sabbatical leave at University of Canterbury in Christchurch, New Zealand. He would like to thank Dr. Daniel Gerhard for his support.

 \bibliographystyle{apalike}
\setlength{\bibsep}{1pt}
\begin{small}

 \end{small}

\section{Appendix: Proof of Theorem \ref{thm7} and Remark \ref{remplac} } \label{sec5}
\def\theequation{5.\arabic{equation}}
\setcounter{equation}{0}

The proof of the theoretical results of this paper proceeds in several steps. First  we state 
the assumptions under which  the statements hold (Section \ref{sec61}). Second, we derive the asymptotic distribution
of the parameter estimates in models with common parameters (Section \ref{sec62}).
In Section  \ref{sec63} we derive a result on the weak convergence of a stochastic process, from which the proof of Theorem \ref{thm7} and Remark \ref{remplac} can be derived
(see Section \ref{sec64}).

\subsection{Assumptions} \label{sec61}

For the theoretical results of this paper we make the same assumptions as in \cite{detmolvolbre2015}. 
\begin{enumerate}
	\item
	The errors $\eta_{\ell,i,j}$ are independent, have finite variance $\sigma_\ell^2$ and expectation zero.
	\item 
	The covariate region $\mathcal{D} \subset \mathbb{R}^d$ is compact and the number and location of dose levels $k_\ell$ does not depend on $n_\ell,\ \ell=1,2$.
	\item
	All estimates of the parameters $\beta_1, \beta_2$ are computed over compact sets $B_1 \subset \mathbb{R}^{p_1}$ and $B_2 \subset \mathbb{R}^{p_2}$.
	\item 
	The regression functions $m_1$ and $m_2$ are twice continuously differentiable with respect to the parameters for all $b_1,b_2$ in neighbourhoods of the true parameters $\beta_1,\beta_2$ and all $d \in \mathcal{D}$. The functions $(d,b_\ell) \mapsto m_\ell(d,b_\ell)$ and their first two derivatives are continuous on $\mathcal{D} \times B_\ell$.
	\item 
	The gradients with respect to the parameters are uniformly bounded, that is \\
	$\sup_{d\in \mathcal{D}}\left\|\tfrac{\partial }{\partial\beta_\ell}  m_\ell(d,\beta_\ell)\right\|<\infty,\ \ell=1,2.$
	\item 
	Defining
	$\psi_{a,\ell}^{(n)}(b) : = \sum_{i=1}^{k_\ell} \frac{n_{\ell,i}}{n_\ell} (m_\ell(d_{\ell,i},a) - m_\ell(d_{\ell,i},b))^2,
	$
	we assume that for any $u>0$ there exists a constant $v_{u,\ell}>0$ such that
	$$
	\liminf_{n_1,n_2 \to \infty} \inf_{a \in B_\ell}\inf_{|b-a| \geq u} \psi_{a,\ell}^{(n)}(b) \geq v_{u,\ell} \qquad \ell=1,2.
	$$
\end{enumerate}

\subsection{Asymptotic properties of the OLS }\label{sec62}

In this section we derive the asymptotic normality of the parameter estimates in models with common parameters.
Observing the definition of the OLS $\hat \beta= (\hat \beta_0,\hat{\tilde \beta}_1,\hat{\tilde \beta}_2)$ in \eqref{ols}  we obtain, by taking the partial derivatives, $\hat \beta$  by the necessary conditions
\begin{align} \nonumber 
0 &=&\sum_{\ell=1}^2\sum_{i=1}^{k_\ell}\sum_{j=1}^{n_{\ell,i}}\big(Y_{\ell,i,j}-m_\ell(d_{\ell,i},(\hat\beta_0,\hat{\tilde \beta}_\ell))\big)\big(\tfrac{\partial }{\partial \beta_0}  m_\ell(d_{\ell,i},\beta_0,\tilde \beta_\ell)\big|_{(\beta_0,\tilde \beta_\ell) = (\hat \beta_0,\hat{ \tilde \beta}_\ell)}\big)^T\\
\label{equ_common}0 &=&
\sum_{i=1}^{k_\ell}\sum_{j=1}^{n_{\ell,i}}\big(Y_{\ell,i,j}-m_\ell(d_{\ell,i},(\hat\beta_0,\hat{\tilde \beta}_\ell))\big)\big(\tfrac{\partial }{\partial \tilde \beta_\ell}  m_\ell(d_{\ell,i},\beta_0,\tilde \beta_\ell)\big|_{(\beta_0,\tilde \beta_\ell) = (\hat \beta_0,\hat{ \tilde \beta}_\ell)}\big)^T,\ \ell=1,2.\end{align}
Defining
\begin{align*} 
h_\ell(d_{\ell,i}) &= \big(\tfrac{\partial}{\partial\beta_0} m_\ell(d_{\ell,i},(\beta_0,\tilde \beta_\ell))\big)^T~, ~~
\tilde{h_\ell}(d_{\ell,i})=\big(\tfrac{\partial}{\partial \tilde \beta_\ell} m_\ell(d_{\ell,i},(\beta_0,\tilde \beta_\ell))\big)^T  \\
\hat h_\ell(d_{\ell,i})&=
\big(\tfrac{\partial }{\partial \beta_0}  m_\ell(d_{\ell,i},\beta_0,\tilde \beta_\ell)\big|_{(\beta_0,\tilde \beta_\ell) = (\hat \beta_0,\hat{ \tilde \beta}_\ell)}\big)^T,\ \hat{ \tilde{h_\ell}}(d_{\ell,i})=\big(\tfrac{\partial }{\partial \tilde \beta_\ell}  m_\ell(d_{\ell,i},\beta_0,\tilde \beta_\ell)\big|_{(\beta_0,\tilde \beta_\ell) = (\hat \beta_0,\hat{ \tilde \beta}_\ell)}\big)^T, 
\end{align*} 
for $\ell=1,2$ we can write
\begin{align*} 
\big(\tfrac{\partial }{\partial \beta}  m_1(d_{1,i},\beta_0,\tilde \beta_1)\big|_{(\beta_0,\tilde \beta_1) = (\hat \beta_0,\hat{ \tilde \beta}_1)}\big)^T=\big(\hat h_1(d_{1,i}),\hat{\tilde{h_1}}(d_{1,i}),0_{p-p'}\big) \\
\big(\tfrac{\partial }{\partial \beta}  m_2(d_{2,i},\beta_0,\tilde \beta_2)\big|_{(\beta_0,\tilde \beta_2) = (\hat \beta_0,\hat{ \tilde \beta}_2)}\big)^T=\big(\hat h_2(d_{2,i}),0_{p-p'},\hat{\tilde{h_2}}(d_{2,i})\big),
\end{align*} 
where $0=0_{p-p'}$ denotes the zero vector in $\mathbb{R}^{p-p'}$. Therefore the equations in \eqref{equ_common} can be summarized to
\be 
0&=&\sum_{i=1}^{k_1}\sum_{j=1}^{n_{1,i}}\big(Y_{1,i,j}-m_1(d_{1,i},(\hat\beta_0,\hat{\tilde \beta}_1))\big)\big(\hat h_1(d_{1,i}),\hat{\tilde{h_1}}(d_{1,i}),0_{p-p'}\big)^T\nonumber\\
&+&\sum_{i=1}^{k_2}\sum_{j=1}^{n_{2,i}}\big(Y_{2,i,j}-m_2(d_{2,i},(\hat\beta_0,\hat{\tilde \beta}_2))\big)\big(\hat h_2(d_{2,i}),0_{p-p'},\hat{\tilde{h_2}}(d_{2,i})\big)^T\nonumber\\
&=&\sum_{\ell=1}^2\sum_{i=1}^{k_\ell}\sum_{j=1}^{n_{\ell,i}}\big(Y_{\ell,i,j}-m_\ell(d_{\ell,i},(\hat\beta_0,\hat{\tilde \beta}_\ell))\big)\big(\tfrac{\partial }{\partial \beta}  
m_\ell(d_{\ell,i},\beta_0,\tilde \beta_\ell)\big|_{(\beta_0,\tilde \beta_\ell) = (\hat \beta_0,\hat{ \tilde \beta}_\ell)}\big)^T.
\ee
A Taylor expansion now yields
\be
0&=&\sqrt{n}(\hat \beta-\beta)\tfrac{1}{n}\sum_{\ell=1}^2\sum_{i=1}^{k_\ell}\sum_{j=1}^{n_{\ell,i}}\big(\tfrac{\partial^2 m_\ell(d_{\ell,i},(\beta_0,\tilde \beta_\ell))}{\partial^2\beta}\big)^T\eta_{\ell,i,j}-\tfrac{\partial m_\ell(d_{\ell,i},(\beta_0,\tilde \beta_\ell))}{\partial\beta}\big(\tfrac{\partial m_\ell(d_{\ell,i},(\beta_0,\tilde \beta_\ell))}{\partial\beta}\big)^T\nonumber\\
&+&\tfrac{1}{\sqrt{n}}\sum_{\ell=1}^2\sum_{i=1}^{k_\ell}\sum_{j=1}^{n_{\ell,i}}\eta_{\ell,i,j}\big(\tfrac{\partial m_\ell(d_{\ell,i},(\beta_0,\tilde \beta_\ell))}{\partial\beta}\big)^T+o_{\mathbb{P}}(1),
\nonumber
\ee
and therefore $\hat \beta$ can be linearized as
\small{\begin{align}\label{linear_common}
	\sqrt{n}(\hat \beta-\beta)&=\tfrac{1}{\sqrt{n}}\sum_{\ell=1}^2\sum_{i=1}^{k_\ell}\sum_{j=1}^{n_{\ell,i}}\eta_{\ell,i,j}\big(\tfrac{\partial m_\ell(d_{\ell,i},(\beta_0,\tilde \beta_\ell))}{\partial\beta}\big)^T\\
	&\cdot\Big(-\tfrac{1}{n}\sum_{\ell=1}^2\sum_{i=1}^{k_\ell}\sum_{j=1}^{n_{\ell,i}}\big(\tfrac{\partial^2 m_\ell(d_{\ell,i},(\beta_0,\tilde \beta_\ell))}{\partial^2\beta}\big)^T\eta_{\ell,i,j}-\tfrac{\partial m_\ell(d_{\ell,i},(\beta_0,\tilde \beta_\ell))}{\partial\beta}\big(\tfrac{\partial m_\ell(d_{\ell,i},(\beta_0,\tilde \beta_\ell))}{\partial\beta}\big)^T\Big)^{-1}+o_{\mathbb{P}}(1).\nonumber
	\end{align}}
\normalsize
Due to the strong law of large numbers it holds
\be\label{stronglaw_common}
&&-\tfrac{1}{n}\sum_{\ell=1}^2\sum_{i=1}^{k_\ell}\sum_{j=1}^{n_{\ell,i}}\big(\tfrac{\partial^2 m_\ell(d_{\ell,i},(\beta_0,\tilde \beta_\ell))}{\partial^2\beta}\big)^T\eta_{\ell,i,j}-\tfrac{\partial m_\ell(d_{\ell,i},(\beta_0,\tilde \beta_\ell))}{\partial\beta}\big(\tfrac{\partial m_\ell(d_{\ell,i},(\beta_0,\tilde \beta_\ell))}{\partial\beta}\big)^T\nonumber\\
&\stackrel{a.s}\longrightarrow& \tfrac{1}{\lambda} \sum_{i=1}^{k_1}\zeta_{1,i}\tfrac{\partial m_1(d_{1,i},(\beta_0,\tilde \beta_1))}{\partial\beta}\big(\tfrac{\partial m_1(d_{1,i},(\beta_0,\tilde \beta_1))}{\partial\beta}\big)^T+\tfrac{\lambda-1}{\lambda} \sum_{i=1}^{k_2}\zeta_{2,i}\tfrac{\partial m_2(d_{2,i},(\beta_0,\tilde \beta_2))}{\partial\beta}\big(\tfrac{\partial m_2(d_{2,i},(\beta_0,\tilde \beta_2))}{\partial\beta}\big)^T\nonumber\\
&=&\tilde \Sigma_1+\tilde \Sigma_2=:\tilde \Sigma,
\ee
where the matrices $\tilde \Sigma_1$ and $\tilde \Sigma_2$ are defined by 
\begin{align*} 
\tilde \Sigma_1 &=
\tfrac{1}{\lambda} \sum_{i=1}^{k_1}\zeta_{1,i}\big(h_1(d_{1,i}),\tilde{h_1}(d_{1,i}),0_{p-p'}\big)^T\big(h_1(d_{1,i}),\tilde{h_1}(d_{1,i}),0_{p-p'}\big) \\
\tilde \Sigma_2& =\tfrac{\lambda-1}{\lambda} \sum_{i=1}^{k_2}\zeta_{2,i}\big(h_2(d_{2,i}),0_{p-p'},\tilde{h_2}(d_{2,i})\big)^T\big(h_2(d_{2,i}),0_{p-p'},\tilde{h_2}(d_{2,i})\big),
\end{align*} 
respectively.
Therefore, using the representation in \eqref{linear_common} and the result in \eqref{stronglaw_common}  we obtain  the asymptotic normality of the estimate, that is
\begin{equation} \label{convergence_common}
\sqrt{n}(\hat \beta-\beta)\stackrel{\mathcal{D}}{\rightarrow}\mathcal{N}(0,\tilde\Sigma^{-1}(\sigma_1^2\tilde \Sigma_1+\sigma_2^2\tilde \Sigma_2)\tilde \Sigma^{-1}).
\end{equation}

\subsection{Weak convergence of a stochastic process } \label{sec63}

The essential step in the proof is a result regarding the weak convergence of the process
$$ \tilde p_n(d)=m_1(d,(\hat \beta_0,\hat{ \tilde \beta}_1))-m_2(d,(\hat \beta_0,\hat{ \tilde \beta}_2))-(m_1(d,(\beta_0,\tilde \beta_1))-m_2(d,(\beta_0,\tilde \beta_2))).$$
We further define
$\tilde \Delta(d,\beta)=\tilde\Delta(d,\beta_0,\tilde \beta_1,\tilde \beta_2)= \tilde\Delta(d,\beta_1,\beta_2)=m_1(d,\beta_1)-m_2(d,\beta_2),$
which yields
$ \tilde p_n(d)= \tilde \Delta(d,\hat \beta)-\tilde \Delta(d,\beta).$


\begin{proof}
	At first we derive a linearization of $\tilde p_n$ by using the Taylor expansion
	$
	\tilde \Delta(d,\hat \beta)=\tilde \Delta(d,\beta)+\big(\tfrac{\partial \tilde \Delta(d,\beta)}{\partial \beta}\big)^T(\hat \beta-\beta)^T+R(\beta),
	$
	where $R(\beta)$ is a remainder term and due to \eqref{linear_common} we have $R(\beta)=O_{\mathbb{P}}(\tfrac{1}{n})$. Therefore it holds
	$
	\sqrt{n} \tilde p_n(d)=\sqrt{n}\big(\tfrac{\partial \tilde \Delta(d,\beta)}{\partial \beta}\big)^T(\hat \beta-\beta)^T+o_{\mathbb{P}}(1)
	$
	uniformly with respect to $d\in \mathcal{D}$ (note that due to the assumptions 
	$\tfrac{\partial \tilde \Delta(d,\beta)}{\partial \beta}$ is continuous in $d\in \mathcal{D}$ and $\mathcal{D}$ is a compact set)
	and with the result from \eqref{linear_common} this can be written as
	\be
	\sqrt{n}\tilde p_n(x)&=&\tfrac{\partial \tilde \Delta(x,\beta)}{\partial \beta}\tfrac{1}{\sqrt{n}}\sum_{\ell=1}^2\sum_{i=1}^{k_\ell}\sum_{j=1}^{n_{\ell,i}}\eta_{\ell,i,j}\tfrac{\partial m(x_{\ell,i},(\beta_0,\tilde \beta_\ell))}{\partial\beta}\tilde \Sigma^{-1}+o_{\mathbb{P}}(1).
	\ee
	Observing  \eqref{linear_common} we define
	\be
	\tilde G_n(d)=\big(\tfrac{\partial \tilde \Delta(d,\beta)}{\partial \beta}\big)^T(\hat \beta-\beta)^T=\big(\tfrac{\partial \tilde \Delta(d,\beta)}{\partial \beta}\big)^T\Big(\tfrac{1}{n}\sum_{\ell=1}^2\sum_{i=1}^{k_\ell}\sum_{j=1}^{n_{\ell,i}}\eta_{\ell,i,j}\tfrac{\partial m_\ell(d_{\ell,i},(\beta_0,\tilde \beta_\ell))}{\partial\beta}\tilde \Sigma^{-1}\Big)^T,
	\ee
	and   obtain
	\be\label{konv_common}
	\sqrt{n}(\tilde p_n(d)-\tilde G_n(d))=o_{\mathbb{P}}(1).
	\ee
	We now define
	$f(d):=\big(\tfrac{\partial \tilde \Delta(d,\beta)}{\partial \beta}\big)^T \Sigma^{-1/2},$
	and we consider the map
	$$
	\tilde \Phi:\left\{\begin{array}{ll} \mathbb{R}^{p_1+p_2-p'}\rightarrow \ell_{\infty}(\mathcal{D})\\
	d\mapsto \left(\Phi(d):d\mapsto f(d)\cdot d \right)\end{array}\right.
	$$
	Obviously, using the same arguments as before, $\tfrac{\partial \tilde \Delta(d,\beta)}{\partial \beta}$ is continuous
	and therefore the same holds for $\tilde \Phi$. Consequently, \eqref{convergence_common}  and the Continuous Mapping Theorem (see for example \cite{vaart2000}[p.7 f.]) yield
	$$
	\big \{  \sqrt{n}\tilde G_n(d)\big \}_{d\in \mathcal{D}} = \big \{   \tilde \Phi\big(\sqrt{n} \Sigma^{1/2}(\hat \beta-\beta)^T\big)(d) \big \}_{d\in \mathcal{D}}\stackrel{\mathcal{D}}{\longrightarrow}
	\big \{  f(d) Z\big \}_{d\in \mathcal{D}}\stackrel{\mathcal{D}}{=}\big \{   \tilde G(d)
	\big \}_{d\in \mathcal{D}},$$
	uniformly with respect to $d\in \mathcal{D}$, where $Z\sim\mathcal{N}(0,I_{p_1+p_2-p'})$, which proves the assertion.
\end{proof}

\subsection{Proof of Theorem \ref{thm7} and Remark \ref{remplac} } \label{sec64}

The assertion of Theorem \ref{thm7} now follows by  the same arguments as given in the proof of Theorem~4 in \cite{detmolvolbre2015}. 
In particular it can be shown that 
$$\sqrt{n}(\tilde d_\infty-d_\infty) \dn \mathcal{Z}:= \max \big \{  \max_{d \in \mathcal{E}^+} \tilde G(d),
\max_{d \in \mathcal{E}^-} (-\tilde G(d)) \big \},
$$
where $\mathcal{E}^{\pm}=\{d\in\mathcal{D}|\ m_1(d,\beta_1)-m_2(d,\beta_2)=\pm d_\infty\}$ denote the sets of extremal points, that is those points, where the 
unknown difference $m_{1} -m_{2}$ attains it maximum absolute deviation. 
The details are omitted for the sake of brevity.

Similarly, note that in  the situation of a common placebo group as described in Remark \ref{remplac} the estimates are obtained by minimizing the sum of squares
 in \eqref{olc_plac}.
As $m_{\ell}^0(0,\tilde \beta^0_\ell )=0$, $\ell=1,2$, this function is the same as the one which is 
 obtained allocating the  observations  at placebo arbitrarily to the two groups. More precisely, we can also write the sum of squares in \eqref{olc_plac} as
$$
\sum_{\ell=1}^2\sum_{i=1}^{k_\ell}\sum_{j=1}^{n_{\ell,i}}\big(Y_{\ell,i,j}-(b_{0,1}+\tilde b_{\ell,1}\cdot m_{\ell}^0(d_{\ell ,i},\tilde b^0_\ell ))\big)^2,
$$
where  $Y_{1,1,j} =  Y_{{0,j}}$ ($j=1, \ldots , n_{1,1}$)  and    $Y_{2,1,j} =  Y_{{0,j}}$ ($j=n_{1,1}+1, \ldots , n_{1,1}+n_{2,1} =n_{0}$).
This corresponds to the minimzation of the sum of squares in \eqref{ols}  with a common intercept $b_{0,1}$, and  
consequently this situation can be treated in the same way as considering two different placebo groups with a common intercept $b_{0,1}$.
By  the same arguments  as  given  in Appendix \ref{sec62} the corresponding estimates are asymptotically normal
distributed (if  $n_{1,1}/n \to c_{1}$ and $n_{2,1}/n \to c_{2}$ for some constants $c_{1}, c_2,\in (0,1)$),  and the proof in Section \ref{sec63}  shows  that  Theorem \ref{thm7} 
remains valid in the model with a common placebo group.
  As a consequence we obtain the claim
in Remark  \ref{remplac}

\end{document}